\documentclass[amsfonts, amssymb, amsmath, reprint, showkeys, nofootinbib, twoside, prb]{revtex4-2}
\usepackage[english]{babel}

\usepackage[utf8]{inputenc}
\usepackage[T1]{fontenc}

\usepackage{amsthm}
\usepackage{mathtools}
\usepackage{units}

\usepackage{xcolor}
\usepackage[pdftex]{graphicx}

\usepackage{hyperref}
\hypersetup{colorlinks=true, citecolor=blue, urlcolor=blue, linkcolor=blue}

\newcommand{\bra}[1]{\langle#1\rvert} % Bra
\newcommand{\ket}[1]{\lvert#1\rangle} % Ket
\newcommand{\expect}[1]{ \langle #1 \rangle} % Expectation value

\newcommand{\operator}[1]{\hat{#1}}
\newcommand{\superoperator}[1]{\hat{\hat{#1}}}

\newcommand{\cone}{\mathrm{i}}
\newcommand{\Tau}{\mathrm{T}}

\renewcommand{\vec}[1]{\boldsymbol{#1}}

\begin{document}
\title{Ultrafast spin dynamics: complementing theoretical analyses by quantum-information measures}

\author{Franziska Ziolkowski}
\author{Oliver Busch}
\email[Correspondence email address: ]{oliver.busch@physik.uni-halle.de}
\author{Ingrid Mertig}
\author{Jürgen Henk}
\affiliation{Institut für Physik, Martin Luther University Halle-Wittenberg, 06099 Halle, Germany}

\date{\today} % Leave empty to omit a date

\begin{abstract}
Theoretical analyses of ultrafast spin dynamics commonly address and discuss simulated phenomena by means of observables, whereas in quantum information theory one often utilizes measures of quantum states. In this Paper we report on possible benefits of quantum information measures in simulations of ultrafast spin dynamics. For Co/Cu heterostructures illuminated by femtosecond laser pulses, we discuss the general behaviour of quantum information measures, in particular distances in Hilbert space and degrees of mixing in the density matrix. The measures are in particular sensitive to variations of the polarization of a laser pulse and the sample composition. Moreover, they are closely related to magnetization and number of excited electrons.
\end{abstract}


\maketitle

\section{Introduction}
\label{sec:introduction}
In experiments and theoretical simulations of ultrafast spin dynamics and spin transport, one often compares results for different setups: temperature~\cite{roth2012temperature, schellekens2013investigating}, material composition~\cite{siegrist2019light, willems2020optical, seifert2016efficient}, sample size and geometry~\cite{torosyan2018optimized, qiu2018layer}, details of a laser pulse~\cite{chekhov2021ultrafast, beigang2019efficient,Papaioannou8405588} --- all of which may be varied. The obtained differences are explained in terms of physical processes and quantities; to name but a few, magnetization as well as charge and spin currents. Although successful and well-established, this approach may be complemented by considering the degree of perturbation.

Take, for example, an electronic charge current brought about by a femtosecond laser pulse, for which it is \textit{a priori} not clear whether an increased photon energy causes a stronger perturbation of the quantum state. On the one hand, more energy is transferred into the system -- what may be regarded as a stronger perturbation. On the other hand electrons may be excited into a regime with decreased density of states, leading to less excited states --- what may be interpreted as a weaker perturbed quantum state. We recall that the amount of absorbed energy~\cite{weber2011ultrafast, tows2019tuning, weber2017laser} or the number of excited electrons can be measured and compared to QI measures on equal footing.

Also quantifying the degree of perturbation by means of different observables may lead to contradicting conclusions. For example, on the one hand the degree can be estimated by means of the temperature obtained from approximating time-dependent occupation-number profiles by a Fermi-Dirac distribution~\cite{rhie2003femtosecond,willems2020optical,Johanssen2013}. On the other hand, the amplitude of an emitted THz signal may be utilized for this purpose~\cite{kampfrath2013terahertz, seifert2016efficient}. 

These brief reflections suggest that it would be desirable to quantify the degree of perturbation of a system independently of observables. Suitable measures are provided by quantum information (QI) theory~\cite{Barnett2009,Nielsen2010}. One subject of QI theory is the general study of information-processing capabilities of quantum systems \cite{Bub2020}. For this purpose, the utilized measures address the degree of coherence of a quantum state or the distance of two quantum states in Hilbert space. A quantum state is described by its density matrix $\operator{\rho}(t)$, the latter entering the time-dependent expectation value
\begin{align}
    \expect{O}(t) & = \operatorname{Tr} \left(\operator{\rho}(t) \, \operator{O} \right)
\end{align}
of an observable $O$ (Ref.~\onlinecite{Gleason1957}). This raises immediately a question regarding relations between observables and QI measures in spin dynamics simulations.

In the present work the time evolution of density matrices of inhomogeneous Co/Cu systems has been studied systematically within our computational framework \textsc{evolve} (for details see Ref.~\onlinecite{Toepler2021}). We address four QI measures, two distance and two mixture measures. While fidelity and trace distance scale the deviation of a quantum state with respect to a reference state, purity and von Neumann entropy quantify the degree of disorder or loss of information. In the present analysis, these measures are related to the magnetization and to the number of electrons excited by the laser pulse. We conclude that QI measures are valuable theoretical tools that complement evaluations based solely on observables.

This Paper is organized as follows. In Section~\ref{sec:theoretical-aspects} we introduce the QI measures~(\ref{sec:metrics}), sketch the spin dynamics simulations~(\ref{sec:spin-dynamics-simulations}), and give details of the setup~(\ref{sec:system}). Results of the simulations are presented and discussed in Section~\ref{sec:results-and-discussion}. We address general properties of the measures and their relation to the magnetization and the number of photo-excited electrons~(\ref{sec:general-behaviour}). Various setups are compared in Section~\ref{sec:comparing-systems}. We conclude and give an outlook in Section~\ref{sec:conclusions-and-outlook}.

\section{Theoretical aspects} \label{sec:theoretical-aspects}
\subsection{Measures in quantum information theory} \label{sec:metrics}
A time-dependent normalized density matrix
\begin{align}
    \operator{\rho}(t) & = \sum_{ij} \ket{i} \, p_{ij}(t) \, \bra{j}
    \label{eq:density-matrix}
\end{align}
is in general nondiagonal. An offdiagonal element $p_{ij}(t)$, i.\,e.\ a coherence, describes the evolution of the coherent superposition of the basis states $\ket{i}$ and $\ket{j}$, while a diagonal element $p_{ii}(t)$ is the occupation probability of state~$\ket{i}$ at time~$t$.

We consider two sets of QI measures: fidelity and trace distance as well as purity and von Neumann entropy. Various distinguishability measures are compared in Ref.~\onlinecite{Audenaert2014}.

Both Uhlmann-Jozsa fidelity~\cite{uhlmann1976transition}
\begin{align}
    F(\operator{\rho}(t), \operator{\rho}(t_{0})) & \equiv \left( \operatorname{Tr} \sqrt{\sqrt{\operator{\rho}(t_{0})} \, \operator{\rho}(t) \sqrt{\operator{\rho}(t_{0})}} \right)^{2}
    \label{eq:fidelity}
\end{align}
and trace distance
\begin{align}
    \Tau(\operator{\rho}(t), \operator{\rho}(t_{0})) & \equiv \frac{1}{2} \operatorname{Tr} \sqrt{ (\operator{\rho}(t) - \operator{\rho}(t_{0}))^{\dagger} (\operator{\rho}(t) - \operator{\rho}(t_{0}))}
    \label{eq:trace-distance}
\end{align}
quantify the congruence between $\operator{\rho}(t)$ and a reference density matrix $\operator{\rho}(t_{0})$, that is the grade of discriminability in the time evolution of the system with respect to the initially prepared system. They fulfill the Josza axioms \cite{Jozsa1994}, take values in $[0, 1]$, and are related by
\begin{align}
\begin{split}
1 - \sqrt{F(\operator{\rho}(t), \operator{\rho}(t_{0}))} & < \Tau(\operator{\rho}(t), \operator{\rho}(t_{0})) 
\\ & \leq \sqrt{1 - F(\operator{\rho}(t), \operator{\rho}(t_{0}))}
\end{split}
\label{eq:relation}
\end{align}
to each other. Fidelity and trace distance are sometimes abbreviated as $F(t)$ and $\Tau(t)$, respectively.

Measures for the degree of mixture in the quantum state are the purity 
\begin{align}
    \gamma(t) & \equiv \expect{\operator{\rho}(t)} = \operatorname{Tr} \operator{\rho}(t)^{2}
    \label{eq:purity}
\end{align}
and the von Neumann entropy~\cite{Bengtsson2006}
\begin{align}
    S(t) & \equiv - \expect{\ln\operator{\rho}(t)} = - \operatorname{Tr} \operator{\rho}(t) \ln\operator{\rho}(t).
    \label{eq:vNE}
\end{align}
A pure state is one of the natural orbitals, from which follows $\gamma = 1$ and $S = 0$. A maximally mixed state is characterized by equipartition among the natural orbitals: its diagonal density matrix has occupation probability $p_{ii} = d^{-1}$ ($d$ dimension of the Hilbert space). Hence, $\gamma = d^{-1}$ and $S = \ln d$ in this case (Ref.~\onlinecite{Jaeger2007}). 

Both distance measures and both mixture measures are in a sense opposed to each other: if the fidelity drops, the trace distance increases, and vice versa [Eq.~\eqref{eq:relation}]. The same holds for the purity and the von Neumann entropy.

For studying their dynamics it is advantageous to consult the relative difference
\begin{align}
    \Delta X(t) & \equiv \frac{| X(t) - X(t_{0}) |}{\operatorname{max}\left( | X(t) - X(t_{0}) | \right)}, \quad t \in  [t_{\mathrm{min}}, t_{\mathrm{max}}],
    \label{eq:relative}
\end{align}
of the above measures, in which $X$ is one of $F$, $\Tau$, $\gamma$, and $S$. The maximum is taken within the considered period.

\subsection{Spin dynamics simulations} \label{sec:spin-dynamics-simulations}
In our theoretical approach \textsc{evolve} for ultrafast spin dynamics~\cite{Toepler2021} the one-electron density matrix $\operator{\rho}(t)$ is evolved in time according to the Lindblad equation
\begin{align}
    \cone \hbar \frac{\mathrm{d} \operator{\rho}(t)}{\mathrm{d} t} & = [\operator{H}(t), \operator{\rho}(t)] + \superoperator{L}[\operator{\rho}(t)].
    \label{eq:LnvL}
\end{align}
From $\operator{\rho}(t)$ we calculate spin polarization, magnetization, as well as charge and spin currents. 

The electronic structure of the sample is modeled in a tight-binding approach for a cluster of atoms, whose static Hamiltonian $\operator{H}_{0}$ includes magnetism and spin-orbit coupling. The density matrix, Eq.~\eqref{eq:density-matrix}, is written in terms of its eigenstates $\{ \ket{i} \}$ with energies $\{ \epsilon_{i} \}$.

The electron system is excited by a laser pulse with an electric field
\begin{align}
    \vec{E}(t) & = \left[ \vec{E}_{\mathrm{s}} \cos(\omega t + \varphi_{\mathrm{s}}) + \vec{E}_{\mathrm{p}} \cos(\omega t + \varphi_{\mathrm{p}}) \right] g(t),
    \label{eq:e-field}
\end{align}
that is a coherent superposition of s- and p-polarized partial waves with amplitudes $\vec{E}_{\mathrm{s}}$ and $\vec{E}_{\mathrm{p}}$ as well as phase shifts $\varphi_{\mathrm{s}}$ and $\varphi_{\mathrm{p}}$, respectively. $g(t)$ is a Gaussian envelope of femtosecond width, and $\hbar \omega$ is the photon energy. $\vec{E}(t)$ enters the Hamiltonian via minimal coupling which is implemented in our computer code using the unitary transformation introduced in Ref.~\onlinecite{Savasta1995}. The time-dependent Hamiltonian $\operator{H}(t)$ in Eq.~\eqref{eq:LnvL} then combines $\operator{H}_{0}$ and the coupling to the laser pulse.

The Lindblad superoperator~\cite{lindblad1976generators, pershin2008effective} $\superoperator{L}[\operator{\rho}(t)]$ in Eq.~\eqref{eq:LnvL} accounts for the coupling of the electron system to a bosonic heat bath. It comprises the Lindbladians of all jump operators $\ket{i} \bra{j}$, which are weighted by the Bose-Einstein distribution of the bath for the prescribed temperature~$T$. Energy is transferred from the electron system into the bath, if $\epsilon_{i} < \epsilon_{j}$ for the involved electronic states, and vice versa, if $\epsilon_{i} > \epsilon_{j}$. The Lindbladians account thus for thermalization of the electron system, typically on the timescale of picoseconds, but also reduce the coherences in the density matrix (that is `dephasing'). For details see Ref.~\onlinecite{Toepler2021}.

A typical simulation consists of three stages. Initially, the electron system is thermalized by coupling to the heat bath. The resulting density matrix $\operator{\rho}(t_{0})$ is diagonal and describes a mixed state, whose Fermi-Dirac distributed occupation probabilities, temperature, and chemical potential are in accordance with the bath's temperature. This state is used when evaluating Eqs.~\eqref{eq:fidelity} and~\eqref{eq:trace-distance}. Second, a laser pulse excites the electron system into a nonthermal state, whose density operator $\operator{\rho}(t)$ is nondiagonal. And third, after the laser pulse the system becomes thermalized again and relaxes slowly toward the initial state, caused by a net transfer of energy from the electron system into the heat bath. 

\subsection{Systems and setups} \label{sec:system}
\begin{figure}
    \centering
    \includegraphics[width=0.9\columnwidth]{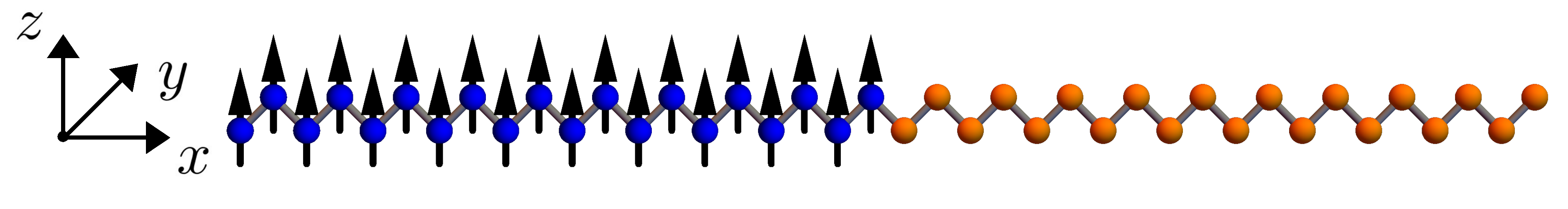}
    \caption{Geometry of the reference system. Co and Cu atoms are arranged in a zig-zag chain along the $x$-axis with 40 sites. Black arrows indicate the magnetic moments oriented along the $z$-axis carried by the Co atoms (blue spheres); the nonmagnetic Cu atoms are shown in orange. The example depicts the case of 20 Co and 20 Cu atoms.}
    \label{fig:geometry}
\end{figure}

For the purpose of this Paper we choose a simple system: a zigzag chain along the $x$-axis whose 40~sites are occupied by either Co or Cu atoms (Fig.~\ref{fig:geometry}). Its electronic structure is described by the tight-binding Hamiltonian $\operator{H}_{0}$ for s-, p-, and d-orbitals of Co and Cu, with Slater-Koster parameters based on those given in Ref.~\onlinecite{Papaconstantopoulos2015}. The magnetic moments of the Co atoms point in $z$ direction (i.\,e.\ perpendicular to the chain). The composition of the system is varied in this study.

The electron system is excited by a femtosecond laser pulse incident within the $xz$-plane with a polar angle of incidence of $\unit[60]{^{\circ}}$ [Eq.~\eqref{eq:e-field}]. The Gaussian envelope is centered at $t = \unit[0]{fs}$, its width is $\unit[10]{fs}$, and its amplitude corresponds to a fluence of about $\unit[300]{mJ/ cm^{2}}$. This value is chosen to produce approximately one excited electron per unit cell. We consider two photon energies ($\unit[0.775]{eV}$ and $\unit[1.55]{eV}$; equivalent to $\unit[0.187]{fs^{-1}}$ and $\unit[0.374]{fs^{-1}}$) as well as pure s and p~polarization (s-polarized: $\vec{E}(t)$ along the $y$ axis; p-polarized: $\vec{E}(t)$ within the $xz$ plane). The temperature of the heat bath is varied as well: $T = \unit[10]{K}$, $\unit[300]{K}$, and $\unit[600]{K}$~(cf.\ Table~\ref{tab:variations}). All other parameters are as in Ref.~\onlinecite{Toepler2021}.

\section{Results and discussion} \label{sec:results-and-discussion}

\subsection{Behaviour of quantum information measures} \label{sec:general-behaviour}
For addressing the general behaviour of the QI measures we choose a Co/Cu chain with 20~sites each (as depicted in Fig.~\ref{fig:geometry}), a p-polarized laser pulse with a $\unit[10]{fs}$ wide envelope centered at $t = \unit[0]{fs}$ and a photon energy of $\unit[1.55]{eV}$. The temperature is $T = \unit[300]{K}$. This is the reference system used in Section~\ref{sec:comparing-systems}.

We first discuss the time dependence of the distance measures fidelity $F$ and trace distance $\Tau$, Eqs.~\eqref{eq:fidelity} and~\eqref{eq:trace-distance}. The reference state $\operator{\rho}(t_{0})$ is the thermalized system at $t_{0} = -\unit[50]{fs}$.

\begin{figure}
    \centering
    \includegraphics[width=0.9\columnwidth]{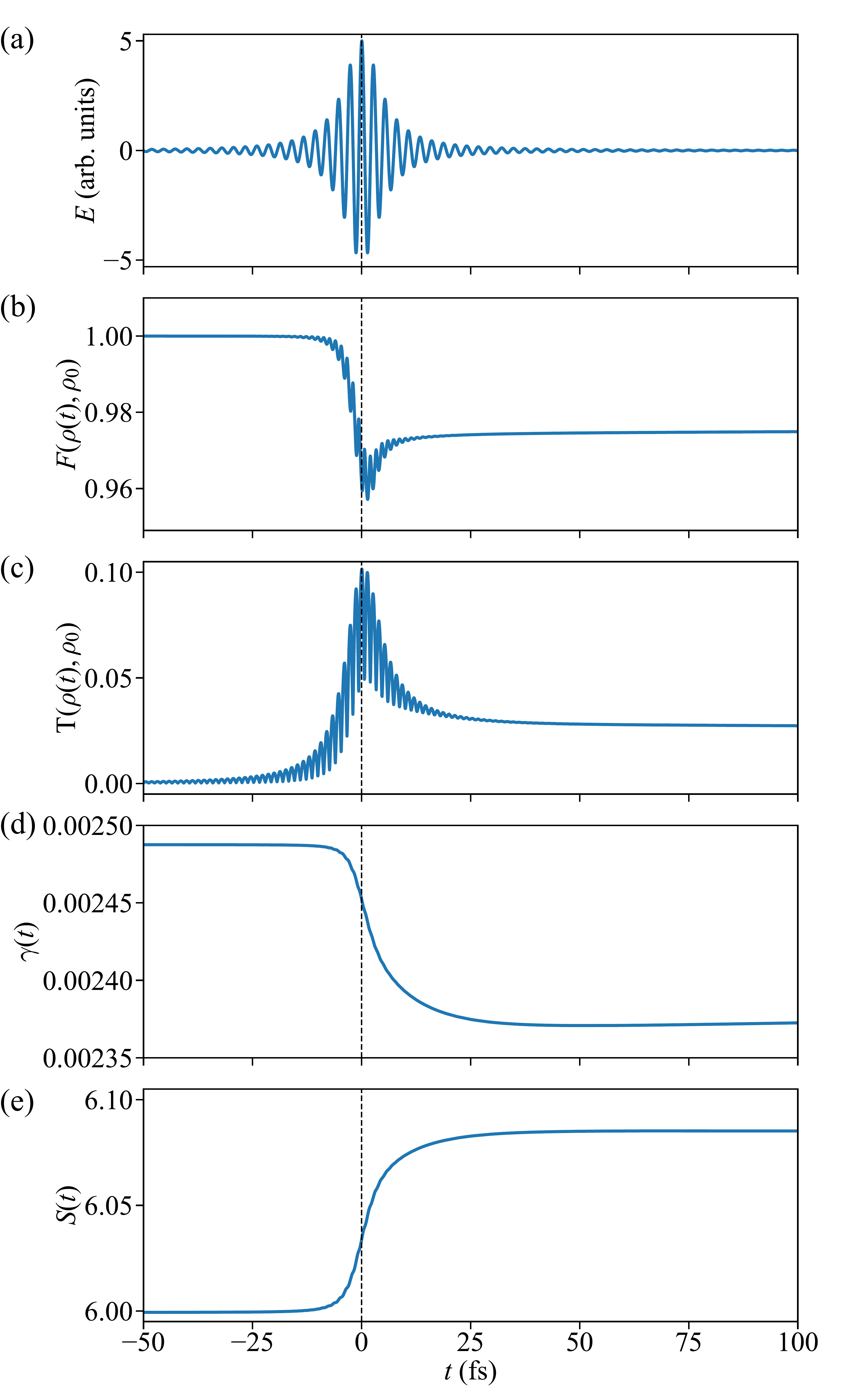}
    \caption{Quantum information measures in the spin dynamics of the Co/Cu chain visualized in Fig.~\ref{fig:geometry}; for details see text. (a) Sketch of the electric field of the laser pulse (in arbitrary units), Eq.~\eqref{eq:e-field}. (b) -- (e) Time dependence of fidelity~$F(t)$ (b), trace distance~$\Tau(t)$ (c), purity~$\gamma(t)$ (d), and von Neumann entropy~$S(t)$ (e), respectively.}
    \label{fig:Chain-general}
\end{figure}

The fidelity $F$ drops by about $\unit[4]{\%}$ (Fig.~\ref{fig:Chain-general}b) caused by the laser pulse (panel~a). Its modulation follows the oscillations of the laser's electric field, but with twice as large a frequency, as has been checked by a Fast-Fourier transformation (FFT) of the signals (confer Fig.~SM1 in the Supplemental Material~\cite{SupplMat}). This behaviour is explained as follows. Dipole transitions cause excitations and de-excitations between the electronic states, irrespectively of the sign of the electric field (cf.\ Fermi's golden rule \cite{Schattke2003}, in which the transition probability is proportional to $\cos^{2}(\omega t) = [ 1 + \cos(2 \, \omega t) ] / 2$). The probability for both processes is largest if the electric field is maximal or minimal. Moreover, the excited electrons propagate within the sample, so that excitation and de-excitation are not reciprocal processes, in particular in inhomogeneous systems. As a result, the fidelity decreases `under the laser pulse'. After the pulse $F$ increases very slowly, caused by the coupling to the heat bath, i.\,e.\ the quantum state relaxes toward the initially thermalized state~$\operator{\rho}(t_{0})$.

The trace distance $\Tau$ displays the same qualitative features as the fidelity (Fig.~\ref{fig:Chain-general}c): modulations and a net increased distance. A closer inspection shows that $\Tau$ `reacts' slightly faster to the laser-induced changes than $F$: compare the onsets of the spectra at $t \approx \unit[-10]{fs}$.

We now turn to the mixture measures purity~$\gamma$ and von Neumann entropy~$S$, Eqs.~\eqref{eq:purity} and~\eqref{eq:vNE}. The thermalization imposes a Fermi-Dirac-type distribution of the occupation probabilities, which yields an initial purity of $\gamma(t_{0}) \approx 0.00249$, a value considerably larger than that of the maximally mixed state with $\gamma = \nicefrac{1}{d} = \nicefrac{1}{720} \approx 0.0014$ (Fig.~\ref{fig:Chain-general}d). The laser pulse introduces coherences~$p_{ij}(t)$ in the density matrix (via dipole transitions) and, thus, decreases the purity (increases the mixing) with a relative drop by about $\unit[5]{\%}$. After the laser pulse the mixing is reduced due to mediation by the heat bath; the purity increases slowly (cf.\ panel d). There is an extremely small modulation with the laser's electric field (not visible in the Figure), in contrast to the sizable ones observed for fidelity and trace distance \cite{FFT}. The von Neumann entropy behaves qualitatively similar to the purity (panel~e). 

Summarizing briefly at this point, a laser pulse changes the electronic quantum state, here in the order of a few per cent, as is quantified by fidelity and trace distance. And it increases the mixing in the same order of magnitude. The two distance measures behave very similarly, as do the mixture measures.

Since the QI measures address the entire system, we compare their time dependencies with that of two observables: magnetization and number of photo-excited electrons. Since we are dealing with inhomogeneous systems, it turns out beneficial to decompose the magnetization of the entire sample into those of the Co and Cu regions.

\begin{figure}
    \centering
    \includegraphics[width=0.9\columnwidth]{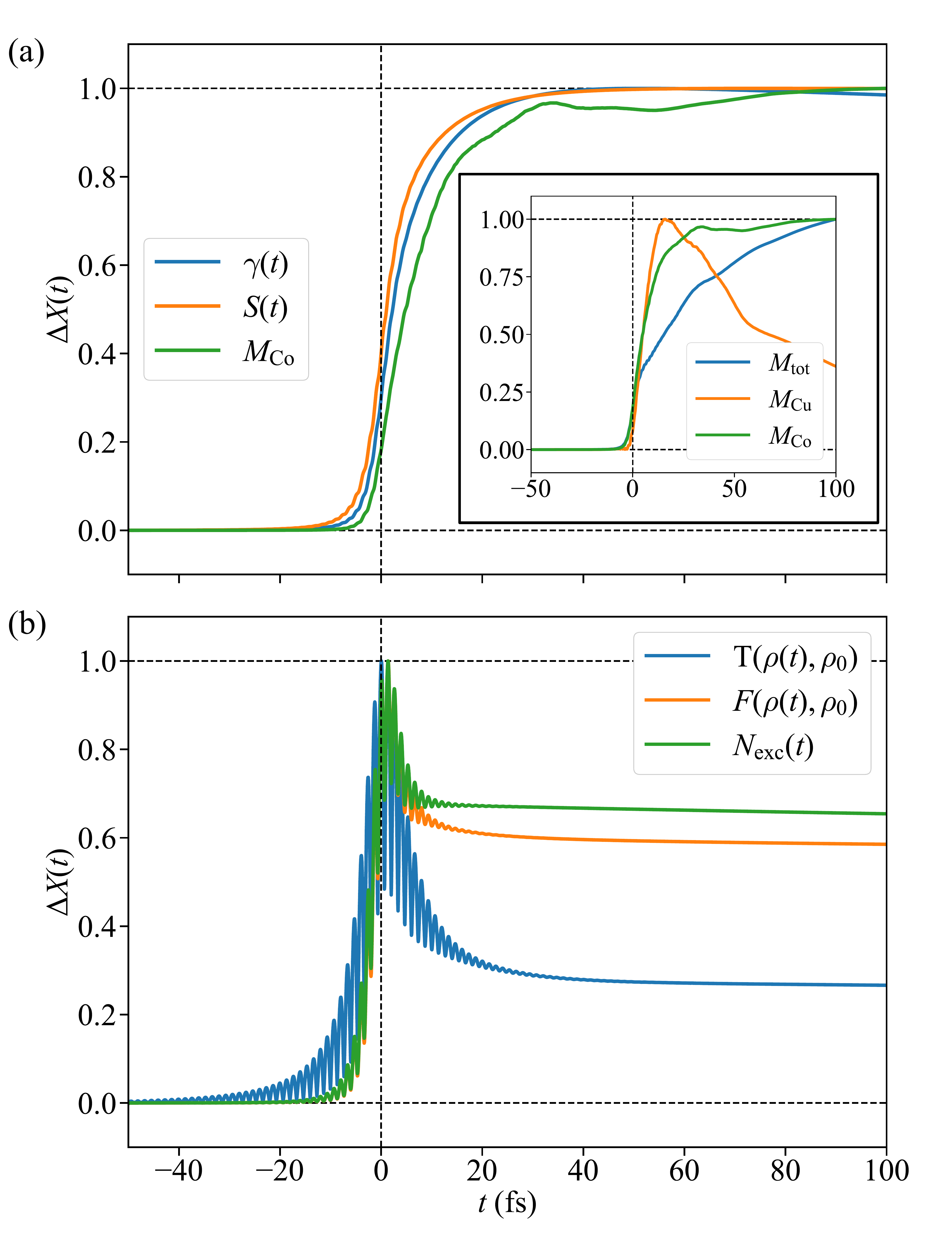}
    \caption{Quantum information measures in the spin dynamics of a Co/Cu chain compared to the magnetization and the number of laser-excited electrons. Data for the measures as in Fig.~\ref{fig:Chain-general}, but displayed as relative changes $\Delta X(t)$, Eq.~\eqref{eq:relative}. (a) Comparison of purity $\gamma(t)$ and entropy $S(t)$ with the magnetization in the Co region of the sample, $M_{\mathrm{Co}}(t)$. The inset shows $\Delta X(t)$ of the  magnetization of the entire sample as well as of its Co and Cu regions, respectively; $M_{\mathrm{tot}}(t) = M_{\mathrm{Co}}(t) + M_{\mathrm{Cu}}(t)$. (b) Comparison of trace distance $\Tau(t)$ and fidelity $F(t)$ with the number of laser-excited electrons $N_{\mathrm{exc}}(t)$.}
    \label{fig:Chain-compare}
\end{figure}

Both purity and entropy follow quite accurately the magnetization change in the Co region (which is actually demagnetized; Fig.~\ref{fig:Chain-compare}a). The laser pulse causes photo-induced spin polarization~\cite{Henk1996, Ebert1996} and affects the coherences in the density matrix. Moreover, spin-orbit coupling leads to spin mixing and, thus, contributes to demagnetization. Both effects increase the mixing in the density matrix.

While the magnetization change builds up quickly in both regions, that in the Co region stays almost constant and that in the Cu region drops rather quickly (Cu becomes magnetized), as is seen in the inset in panel~a. As has been illustrated in Refs.~\onlinecite{Toepler2021} and~\onlinecite{Chen2019}, the interface acts as a `source' of spin-polarized currents in the sample. More precisely, the magnetization change in the Cu region is initiated at the interface by a backflow mechanism~\cite{Chen2019, borchert2020manipulation}; it then spreads out into the Cu region via a spin-polarized current with rapidly decreasing magnitude. As a result, the Cu magnetization exhibits a peak at $t \approx \unit[15]{fs}$, i.\,e.\ slightly after the laser pulse. The interface also affects the demagnetization in the Co region~\cite{Toepler2021}. In addition demagnetization happens at all Co sites, as in a bulk sample. The combination of Cu and Co profile results in the `retarded' profile of the total magnetization.

Another observable is the number of electrons excited by the laser pulse,
\begin{align}
    N_{\mathrm{exc}}(t) & \equiv \sum_{i} \left( p_{ii}(t) - p_{ii}(t_{0}) \right) \, \Theta(\epsilon_{i} - \mu).
    \label{eq:Nexel}
\end{align}
The Heaviside function $\Theta$ selects electronic states with energies $\epsilon_{i}$ larger than the chemical potential $\mu$. To extract the number of laser-excited electrons, the number of thermally excited electrons is subtracted ($p_{ii}(t_{0})$ accounts for the thermally excited electrons described by $\operator{\rho}(t_{0})$). The number of excited electrons is accessible by time-resolved photoelectron spectroscopy~\cite{rhie2003femtosecond,Johanssen2013}, for the number of photoelectrons resembles $N_{\mathrm{exc}}(t)$. Hence, the QI measures may be contrasted with experimental outcomes.

The distance measures follow closely $N_{\mathrm{exc}}(t)$ (Fig.~\ref{fig:Chain-compare}b). All three profiles are modulated by the laser pulse and are peaked at $t = \unit[0]{fs}$, that is at the maximum laser intensity. However, $F(t)$ and $N_{\mathrm{exc}}(t)$ increase more sharply than $\Tau(t)$ at $t < \unit[0]{fs}$. Comparing qualitatively the chronological sequences of panel~a with those in panel~b it is evident that purity and entropy do not resemble $N_{\mathrm{exc}}(t)$. This finding suggests that --- since only diagonal elements $p_{ii}(t)$ enter $N_{\mathrm{exc}}(t)$ --- the distance measures appear largely sensitive to changes of the occupation probabilities, that is to photo-induced excitations and de-excitations. On the contrary, mixture measures reflect magnetization changes.

\subsection{Application: comparing systems} \label{sec:comparing-systems}
In order to compare systems we chose as a reference the one studied in Section~\ref{sec:general-behaviour}: a temperature of $\unit[300]{K}$, a p-polarized laser pulse with $\unit[1.55]{eV}$ photon energy and a sample composition of 20~Co and 20~Cu atoms (short: $\nicefrac{\mathrm{Co}}{\mathrm{Cu}} = \nicefrac{20}{20}$). In each variation a single parameter is modified; see Table~\ref{tab:variations}.

\begin{table}
    \centering
    \begin{tabular}{ll}
    \hline\hline
    \multicolumn{1}{c}{Parameter} & \multicolumn{1}{c}{Variation} \\
    \hline
    Temperature       & $\unit[10]{K}, \mathbf{\unit[300]{K}}, \unit[600]{K}$ \\
    Light polarization & s-polarized, \textbf{p}-polarized \\
    Photon energy & $\unit[0.775]{eV}, \mathbf{\unit[1.55]{eV}}$ \\
    Sample composition $\nicefrac{\mathrm{Co}}{\mathrm{Cu}}$  & $\nicefrac{40}{0}, \nicefrac{30}{10},  \nicefrac{\mathbf{20}}{\mathbf{20}}, \nicefrac{10}{30}, \nicefrac{0}{40}$ \\
    \hline \hline
    \end{tabular}
    \caption{Parameter variations for system comparisons. The parameters of the reference system are typeset in \textbf{boldface}. The sample composition $\nicefrac{\mathrm{Co}}{\mathrm{Cu}}$ is given as ratio of the number of Co and Cu atoms.}
    \label{tab:variations}
\end{table}

\subsubsection{Fidelity} Since trace distance and fidelity behave quite similar, we focus on the fidelity in the following discussion. Results for the trace distance are given in the Supplemental Material~\cite{SupplMat}, see Fig.~SM2a--d.

Temperature enters a simulation in two ways. First, it affects the initial occupation probabilities $p_{ii}(t_{0})$ of the $\operator{H}_{0}$ eigenstates with energies close to the chemical potential (Fermi-Dirac distribution). Second, it enters the Bose-Einstein distribution of the heat bath, which alters the strength of the coupling of the electron system to the bath via the Lindbladians.

\begin{figure}
    \centering
    \includegraphics[width=\columnwidth]{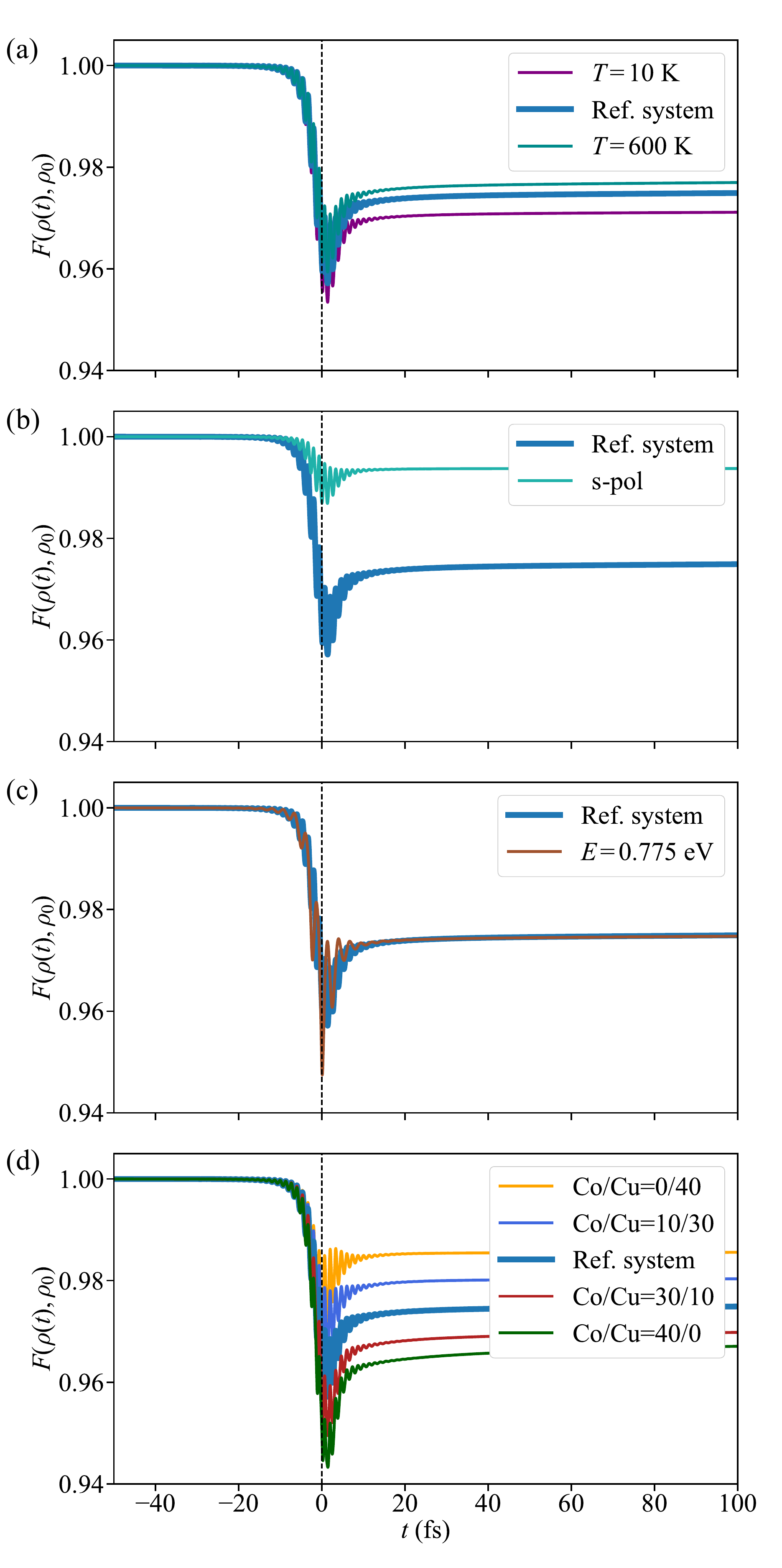}
    \caption{Comparing systems: fidelity. The fidelity $F$ quantifies the variation of systems in which one parameter is modified according to Table~\ref{tab:variations}. Results for varying temperature (a), light polarization (b), photon energy (c), and sample composition (d) are shown.}
    \label{fig:comparison-fidelity}
\end{figure}

The fidelity $F$ shows minute variations with temperature `under the laser pulse' (Fig.~\ref{fig:comparison-fidelity}a). All three spectra exhibit almost identical initial slopes and minima (about $0.96$ at $t = \unit[0]{fs}$). However, a high temperature leads to a faster relaxation after the laser pulse than a low temperature; confer the data for $T = \unit[10]{K}$ and $\unit[600]{K}$. This finding suggests that the initial Fermi-Dirac-type occupation is of minor importance, and the increase after the laser excitation is attributed to the stronger coupling to the heat bath at elevated temperatures.

Contrary to temperature, light polarization has a pronounced effect on the fidelity (Fig.~\ref{fig:comparison-fidelity}b). The minimum for s-polarized light is less deep than for p-polarized light ($0.96$ at $t = \unit[0]{fs}$ compared to $0.99$ at $t = \unit[-50]{fs}$). The dipole selection rules tell which orbitals are involved in the optical transitions. Within the presented geometry, p-polarized light perturbs the system more efficiently than s-polarized light, which suggests to utilize light polarization as means to tailor ultrafast demagnetization and transport phenomena.

The photon energy shows a minute effect on the fidelity (Fig.~\ref{fig:comparison-fidelity}c), although the photon energy $\hbar \omega = \unit[0.775]{eV}$ is only half as large as that of the reference system. This finding is in line with the electronic structures of Co and Cu. The $d$~bands of Co cover energies from about $\unit[-4]{eV}$ up to about $\unit[+1.5]{eV}$, and the $sp$-band `cuts' through this range. $p\leftrightarrow d$ transitions are thus accessible for both selected photon energies. The same holds for $d\rightarrow p$ transitions in Cu. This result is in agreement with experimental outcomes for Fe and Fe/Pt multilayers, which show only minor variations with a change of the excitation wavelength~\cite{chekhov2021ultrafast, beigang2019efficient}.

For investigating the effect of sample composition on the fidelity, the total number of sites is fixed (here: $40$). The largest change is found for pure Co, the smallest for pure Cu (Fig.~\ref{fig:comparison-fidelity}d): the more Co in the system, the larger the change. Cu has a large density of occupied $d$~states and a small density of $sp$~states above the chemical potential, as compared to Co. Hence, optical excitations are less likely in Cu, which is reflected in the smaller decrease of fidelity.

To wrap up, distance measures are mostly sensitive to light polarization and sample composition, rather than to temperature and photon energy.

\subsubsection{Purity} We address the purity now, results for the von Neumann entropy are given in the Supplemental Material~\cite{SupplMat}, see Fig.~SM2e--h.

\begin{figure}
    \centering
    \includegraphics[width=\columnwidth]{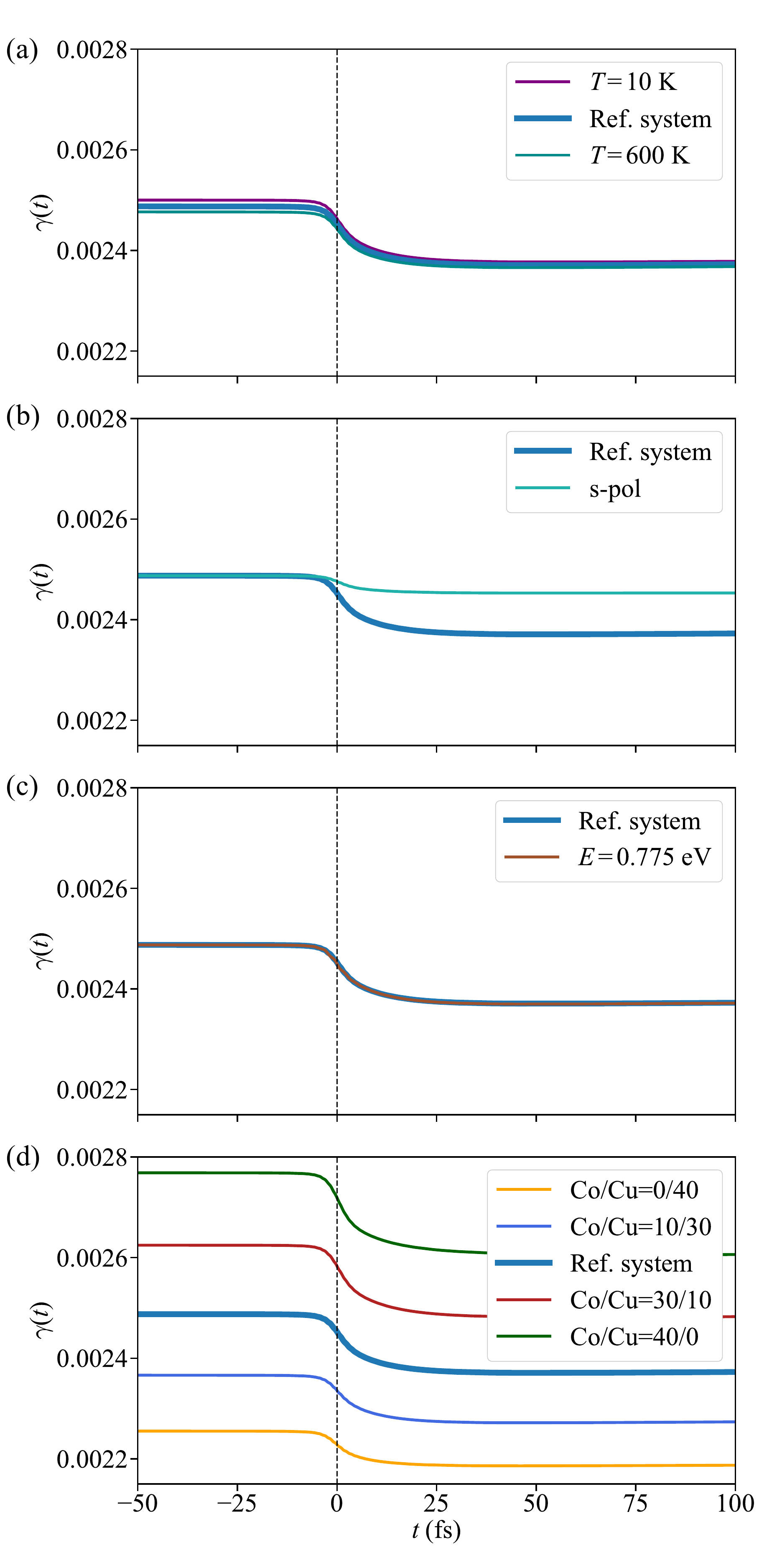}
    \caption{Comparing systems: purity. As Fig.~\ref{fig:comparison-fidelity}, but for the purity~$\gamma$.}
    \label{fig:comparison-purity}
\end{figure}

The effect of the temperature is very small, for only the occupation probabilities of states with energies in a small range about the chemical potential are affected. Consequently, a lower temperature yields marginally less mixing than a higher temperature (Fig.~\ref{fig:comparison-purity}a), as is best seen at times `before the laser pulse'.

As for the fidelity (Fig.~\ref{fig:comparison-fidelity}b), the light polarization variation changes strongly the purity (Fig.~\ref{fig:comparison-purity}b). For both measures, the effect is smaller for s-polarized than for p-polarized light. Moreover, the photon energy shows no visible effect in both measures (cf.\ panel~c).

The sample composition in panel d yields two effects on the purity. 
First, increasing the Cu content increases the number of electrons in the system, which reduces the occupation probabilities of the thermally occupied states. As a result, the purity decreases with increased Cu content. Second, the laser pulse increases the mixing stronger in a pure Co sample than in pure Cu sample, as is evident from the drop of each profile during and after the laser pulse.

The mixing measures are significantly sensitive to light polarization and sample composition, but not to photon energy and temperature -- as is found also for the distance measures. Therefore, explanations valid for the fidelity spectra of Fig.~\ref{fig:comparison-fidelity} apply also to the purity spectra of Fig.~\ref{fig:comparison-purity}.

\subsubsection{Observables: magnetization in the Co region and number of excited electrons} 
\begin{figure}
    \centering
    \includegraphics[width=\columnwidth]{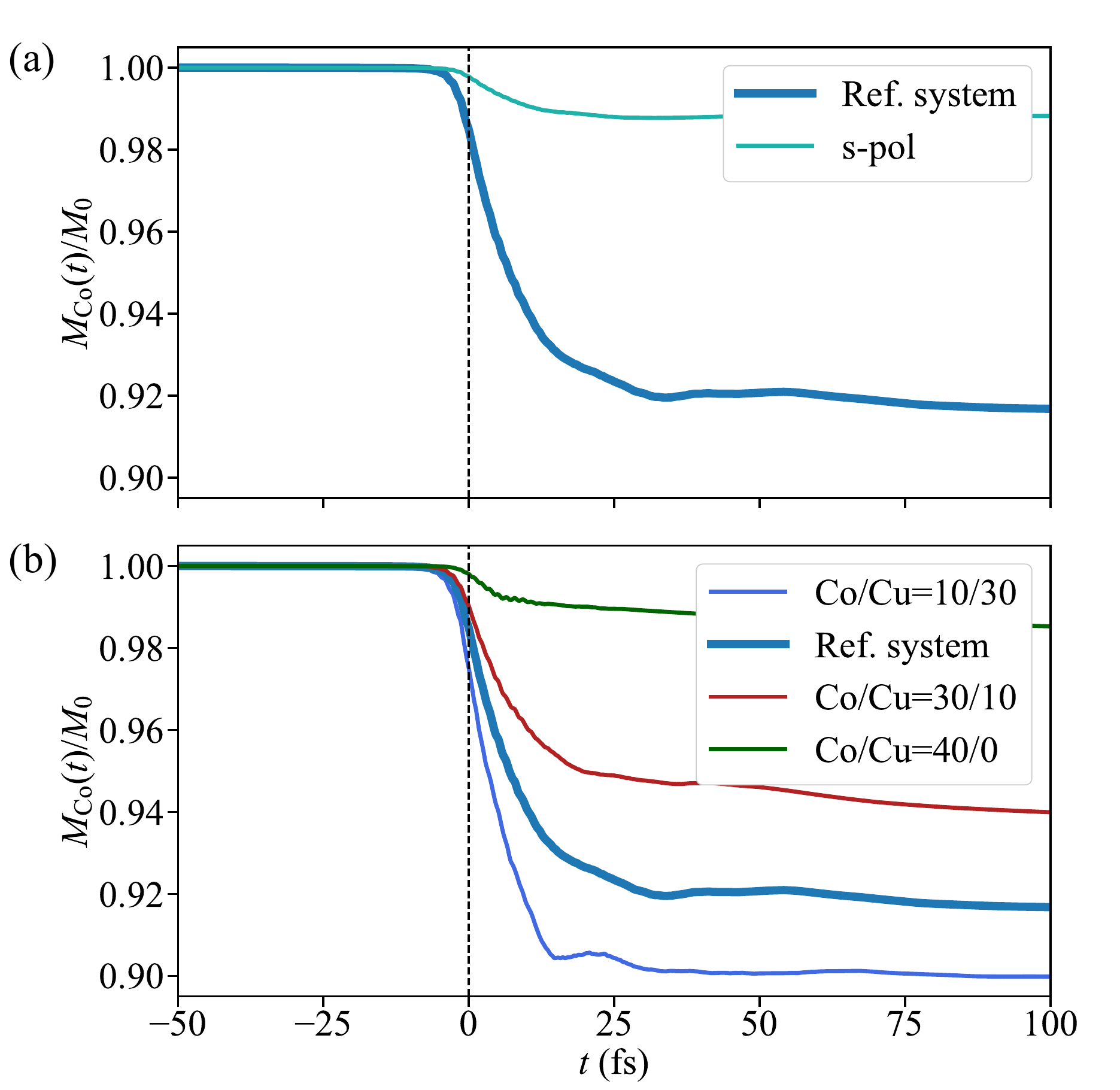}
    \caption{Comparing systems: magnetization $M_{\mathrm{Co}}(t)$ in the Co regions of the samples for variations of the laser polarization~(a) and of the sample composition~(b). Each spectrum is normalized with respect to the magnetization $M_{0}$ at $t = -\unit[50]{fs}$ of the associated system. Data for pure Cu ($\nicefrac{\mathrm{Co}}{\mathrm{Cu}} = \nicefrac{0}{40}$) are not shown.}
    \label{fig:comparison-Co-magnetization}
\end{figure}

In order to complete the discussion, we present data for the magnetization in the Co region of the samples, $M_{\mathrm{Co}}$ in Fig.~\ref{fig:comparison-Co-magnetization} (the magnetization of the entire sample and those in the Cu region are displayed in Fig.~SM3 of the Supplemental Material~\cite{SupplMat}). In Fig.~\ref{fig:comparison-excited_electrons} we show analogous calculations for the number of laser-excited electrons $N_\mathrm{exc}(t)$. We restrict ourselves to discussing the dependencies on light polarization and sample composition, both of which produce pronounced effects in these observables. Variations of temperature and photon energy bring forth minute effects and are depicted in Fig.~SM4 in the Supplemental Material~\cite{SupplMat}.

For s-polarized light, the demagnetization as well as the number of photo-excited electrons are considerably less than for p-polarized light (Figs.~\ref{fig:comparison-Co-magnetization}a and~\ref{fig:comparison-excited_electrons}a) --- as already found for the measures. 

The importance of an interface becomes evident when considering the sample composition (Fig.~\ref{fig:comparison-Co-magnetization}b). While for small Cu content a minor photo-induced spin polarization produces a minute modulation at $t \approx \unit[20]{fs}$ (cf.\ Fig.~SM3h in the Supplemental Material~\cite{SupplMat}) and a small demagnetization for pure Co (green in Fig.~\ref{fig:comparison-Co-magnetization}b, $\nicefrac{\mathrm{Co}}{\mathrm{Cu}} = \nicefrac{40}{0}$), the demagnetization is largest for the system with $\nicefrac{\mathrm{Co}}{\mathrm{Cu}} = \nicefrac{10}{30}$ (blue graph, drop around $t = \unit[0]{fs}$). This finding has been attributed to the imbalance of occupation at the Co/Cu interface (see Ref.~\onlinecite{Toepler2021} for details) which facilitates demagnetization and spin transfer from Cu into Co states, thereby triggering spin currents. This interface-driven effect becomes relatively weaker, the larger the Co content in the sample. It is not revealed in the measures, which implies that QI measures cannot fully replace observables, but complement theoretical analyses based on the latter.

The amount of Cu in the sample strongly affects the number of laser-excited electrons, as well (cf.\ Fig.~\ref{fig:comparison-excited_electrons}b). After the laser pulse, most electrons are in an excited state in a pure Co sample (green graph). With increasing Cu content the amount of excited electrons is decreasing since optical excitations in Cu are less likely as explained in~\ref{sec:comparing-systems}~1. for the fidelity.

\begin{figure}
    \centering
    \includegraphics[width=\columnwidth]{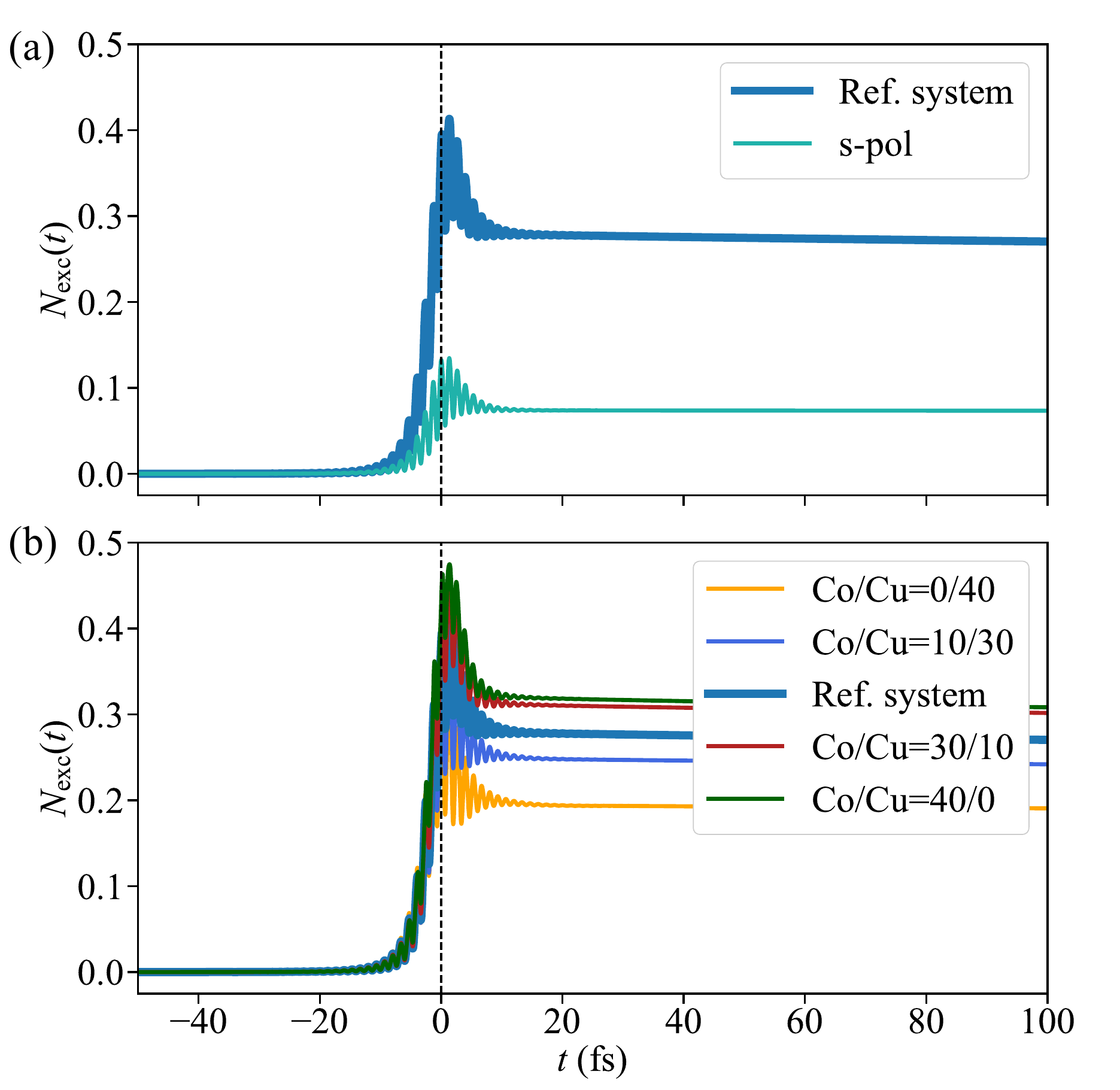}
    \caption{Comparing systems: as Fig.~\ref{fig:comparison-Co-magnetization} but for the number of laser-excited electrons $N_{\mathrm{exc}}(t)$. In contrast to Fig.~\ref{fig:comparison-Co-magnetization}, data for a sample consisting only of Cu ($\nicefrac{\mathrm{Co}}{\mathrm{Cu}} = \nicefrac{0}{40}$) is shown in yellow.}
    \label{fig:comparison-excited_electrons}
\end{figure}

\section{Concluding remarks} \label{sec:conclusions-and-outlook}
The above discussion shows that quantum information measures are suitable tools for theoretical analyses of the quantum state in ultrafast spin dynamics that complement commonly utilized observables. Although sensitive in particular to the light polarization of a laser pulse and to sample composition, the subtle differences of the four measures discussed here suggest to utilize at least one distance measure and one mixture measure to characterize the degrees of perturbation and coherence of the quantum state under the influence of a laser pulse. We would like to stress that the minute extra effort in computing the QI measures during a simulation is worth the extra insight gained. Moreover, we hope that our findings encourage colleagues to utilize QI measures in theoretical spin dynamics investigations.

Concerning future applications we think it worth to compare THz with laser excitations (i.\,e.\ excitations close to the chemical potential versus excitations within a significantly broader energy range), linearly versus circularly polarized light, and sample composition (e.\,g., replacing Co/Cu by Fe/Al).

\begin{acknowledgments}
This work is supported by TRR~227 of Deutsche Forschungsgemeinschaft (project~B04).
\end{acknowledgments} 

\bibliography{references}

\clearpage

\onecolumngrid

\appendix
\section*{Supplemental material}

\section{Frequency analysis} 
\begin{figure}[h]
	\centering
	\includegraphics[width=0.667\columnwidth]{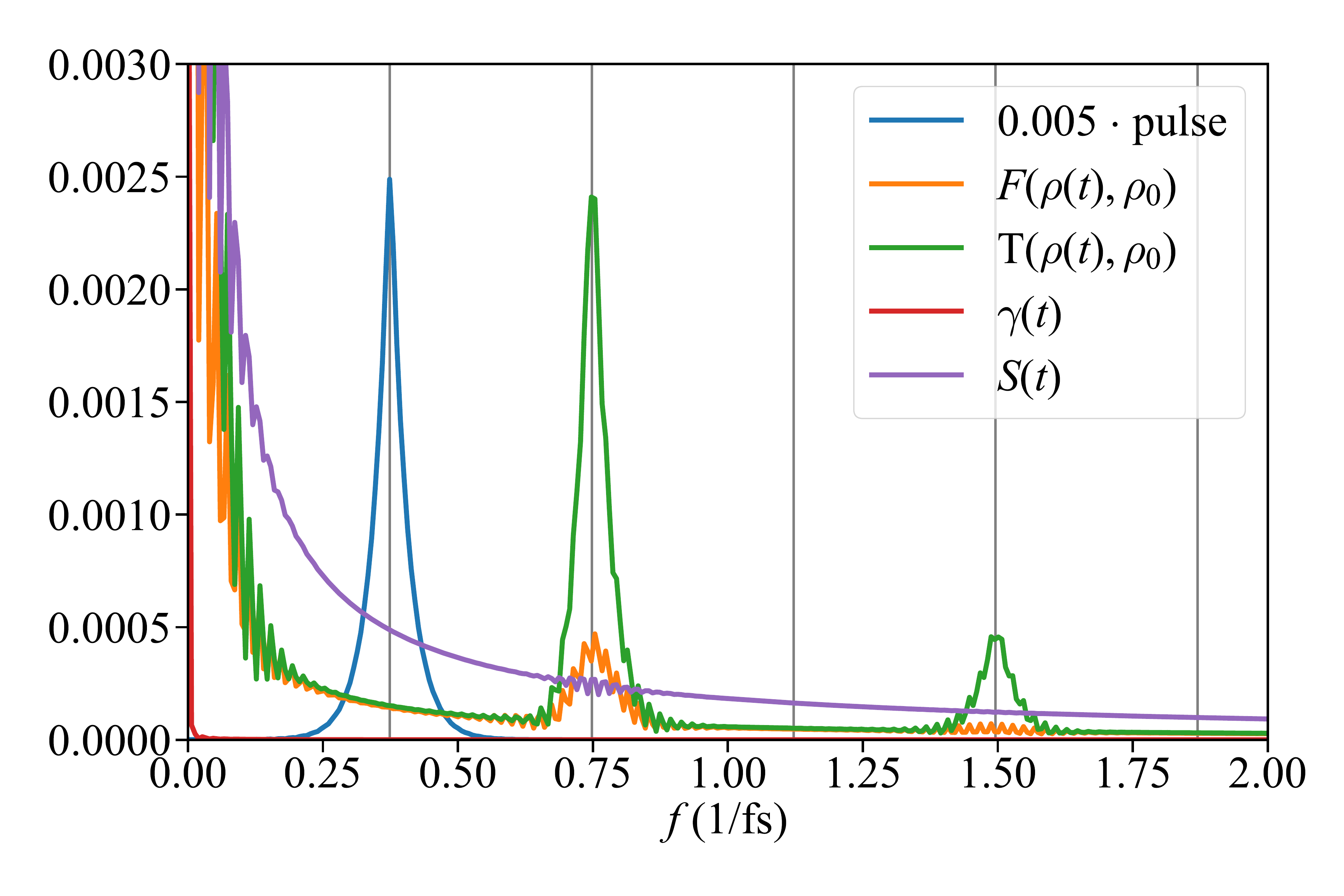}
	\caption{Fast-Fourier transformation of laser-pulse shape (blue), fidelity $F(t)$ (orange), trace distance $\Tau(t)$ (green), purity $\gamma(t)$ (red), and von Neumann entropy $S(t)$ (purple) from Fig.~2 in the main text. The vertical black lines indicate the integer multiples of the laser's frequency $f = \unit[0.374]{fs^{-1}}$.}
	\label{fig:FFT_pulse_and_distances}
\end{figure}

The doubled frequency of fidelity~$F(t)$ and trace distance~$\Tau(t)$ with respect to the laser's frequency (see Figs.~2 and~4 in the main text as well as the left column of Fig.~\ref{fig:comparison-trace-distance-entropy}) has been checked by Fast-Fourier transformations (FFT). The laser's photon energy of $\unit[1.55]{eV}$ is equivalent to a frequency of $f = \unit[0.374]{fs^{-1}}$, resulting in a single peak in the FFT spectrum (blue in Fig.~\ref{fig:FFT_pulse_and_distances}; the broadening is caused by the Gaussian envelope). In contrast, both measures do not exhibit a maximum at that frequency but at the doubled and quadrupled values of $f \approx \unit[0.75]{fs^{-1}}$ and~$\approx \unit[1.50]{fs^{-1}}$ (cf.\ vertical black lines).
Purity~$\gamma(t)$ and von Neumann entropy~$S(t)$ exhibit modulations at about $\unit[0.75]{fs^{-1}}$ that are several orders of magnitude smaller than those of fidelity and trace distance, and thus not resolved in this Figure.

\section{Comparing systems}
\subsection{Trace distance and von Neumann entropy}
\label{app:measures}
\begin{figure*}
	\centering
	\includegraphics[width=0.48\columnwidth]{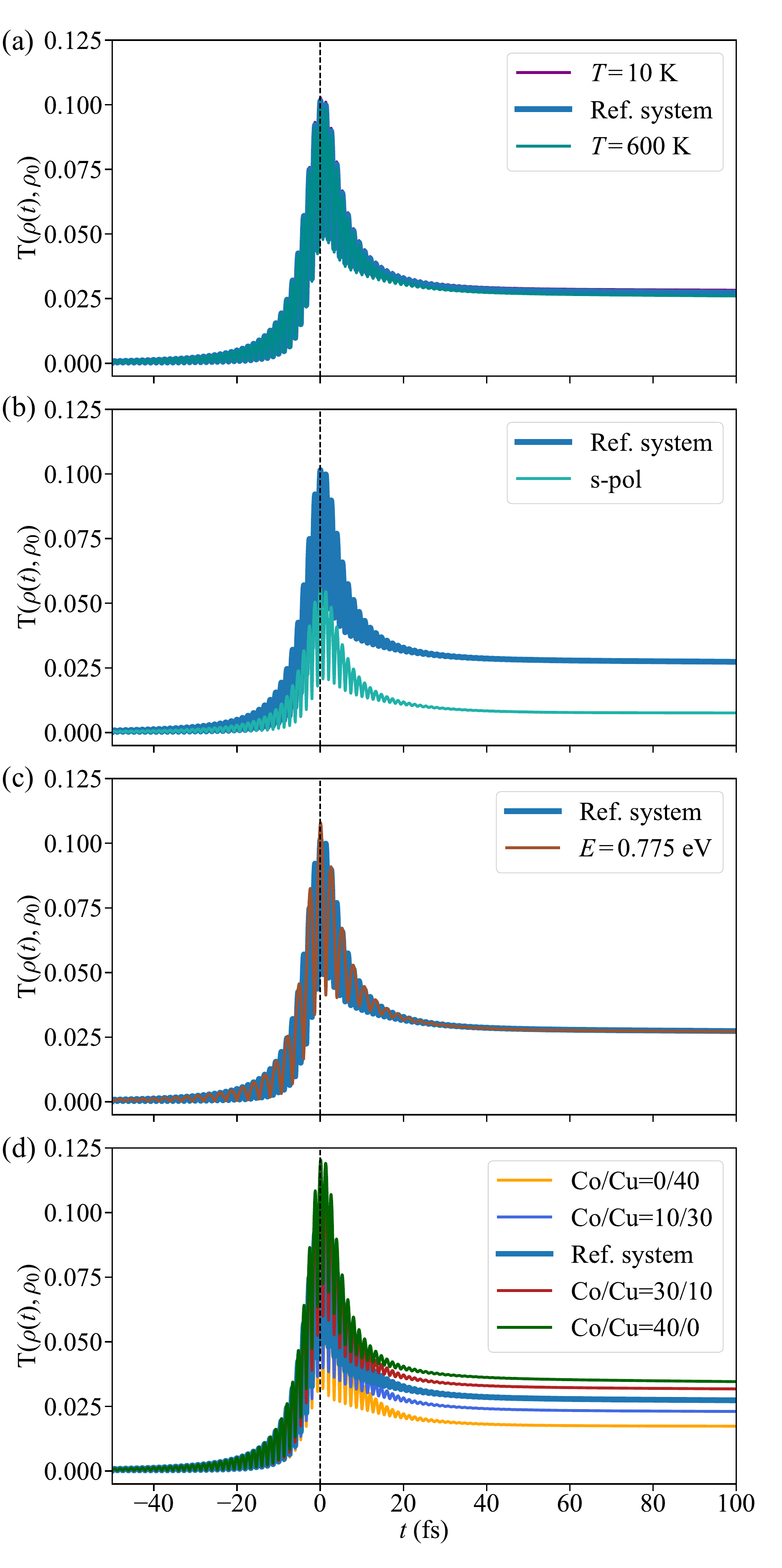}
	\includegraphics[width=0.48\columnwidth]{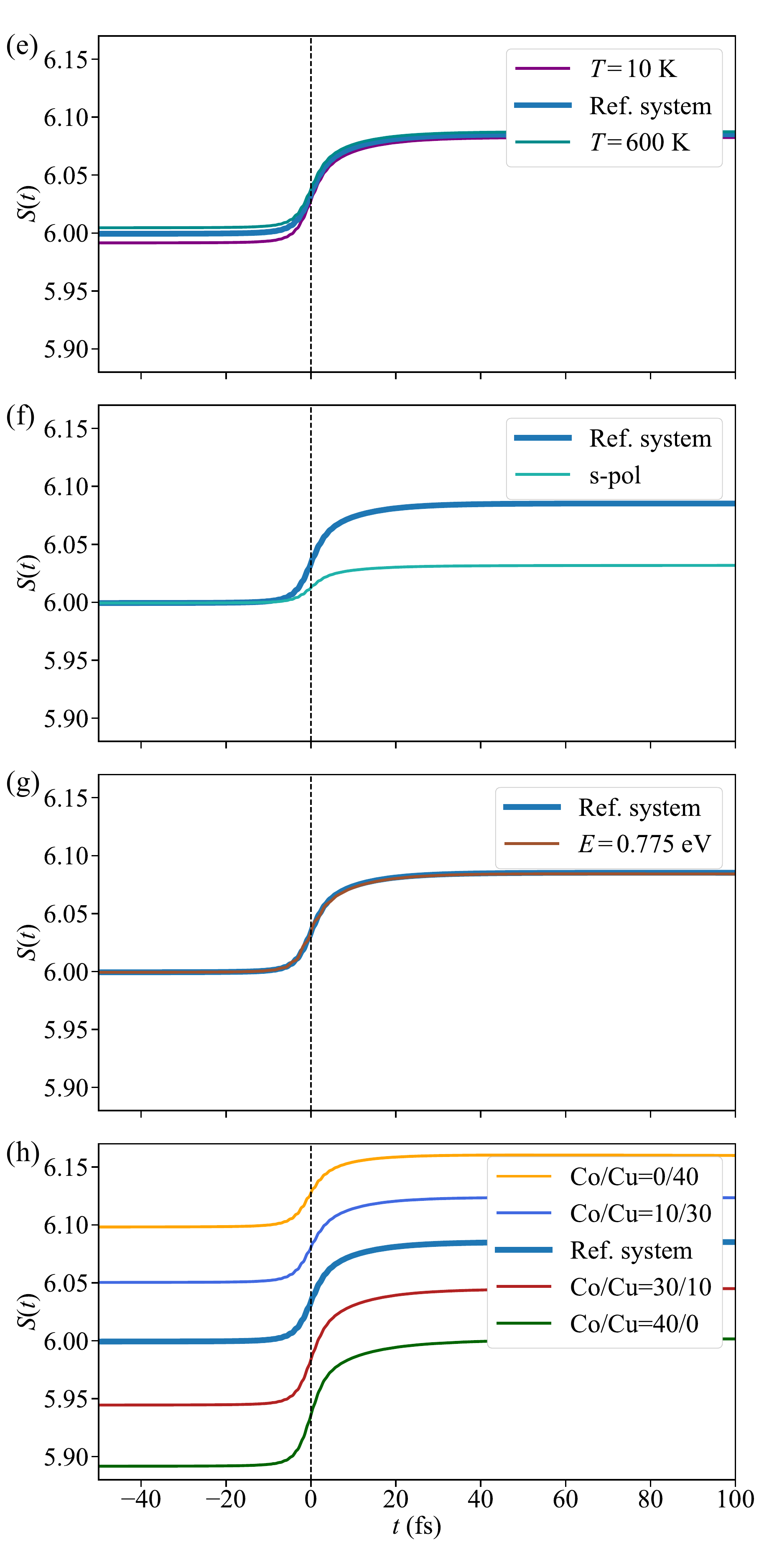}
	\caption{Comparing systems: trace distance~$\Tau(t)$ [left column, panels (a)--(d)] and von Neumann entropy~$S(t)$ [right column, panels (e)--(h)]. Both measures quantify the variation of systems in which one parameter is modified according to Table~1 in the main publication. Results for varying temperature (a) and (e), light polarization (b) and (f), photon energy (c) and (g), and sample composition (d) and (h) are displayed.}
	\label{fig:comparison-trace-distance-entropy}
\end{figure*}

The investigation of the measures revealed that fidelity~$F(t)$ and trace distance~$\Tau(t)$ show qualitatively similar behaviour (cf.\ Figs.~2 and~3 in the main text). The same holds for the two mixture measures purity~$\gamma(t)$ and von Neumann entropy~$S(t)$. To illustrate this in more detail, Figure~\ref{fig:comparison-trace-distance-entropy} displays the trace distance~$\Tau(t)$ [left column, panels~(a)--(d)] and the von Neumann entropy~$S(t)$ [right column, panels~(e)--(h)] for the various systems discussed in Section~III.B of the main text. The variations of temperature and photon energy lead to minimal changes, whereas those for light polarization and sample composition have a strong effect, which is in agreement with our findings for fidelity and purity (cf.\ Figs.~4 and 5~in main text).

\subsection{Total and Cu magnetization}
\label{app:magnetization}
\begin{figure*}
	\centering
	\includegraphics[width=0.48\columnwidth]{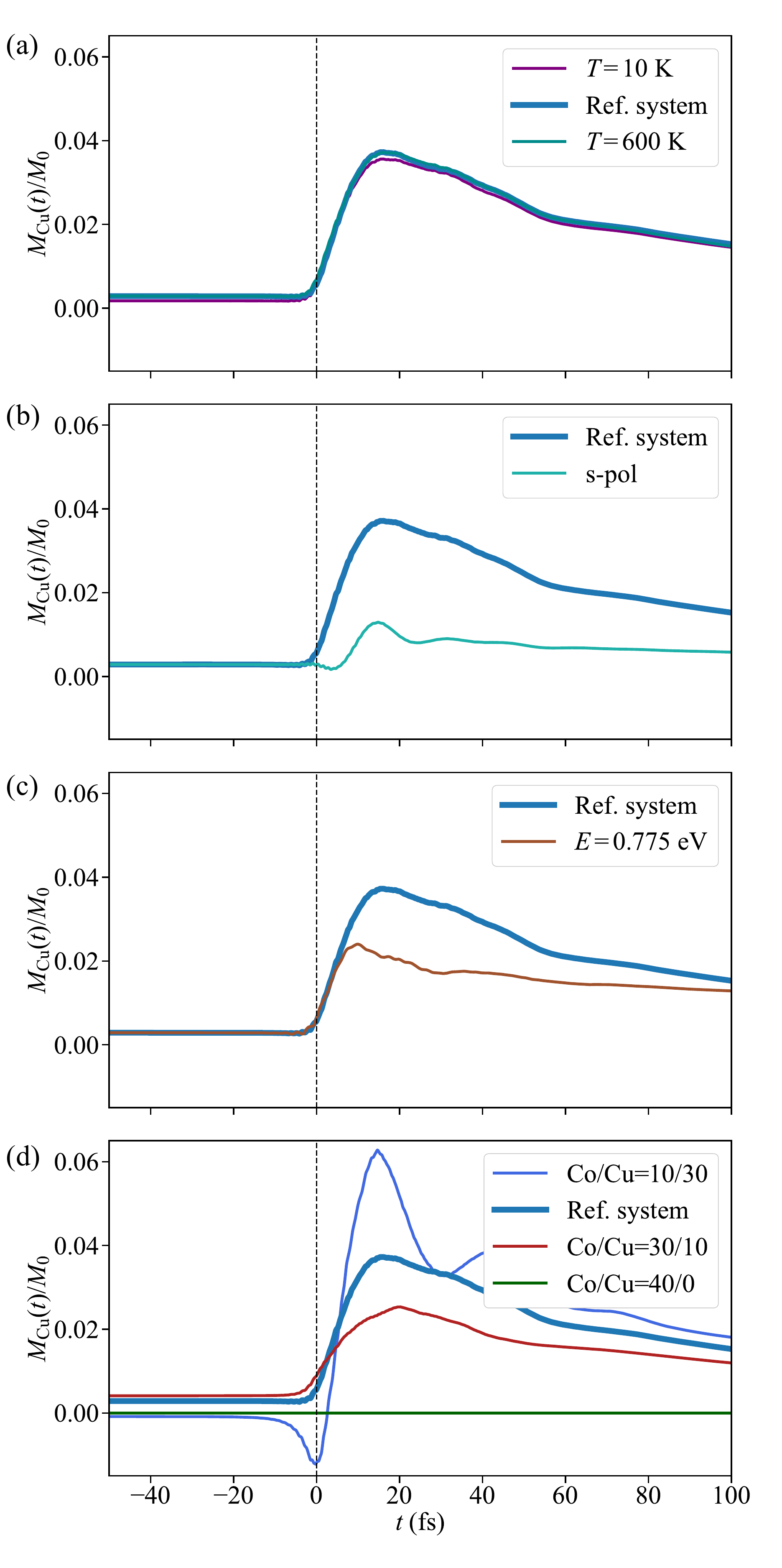}
	\includegraphics[width=0.48\columnwidth]{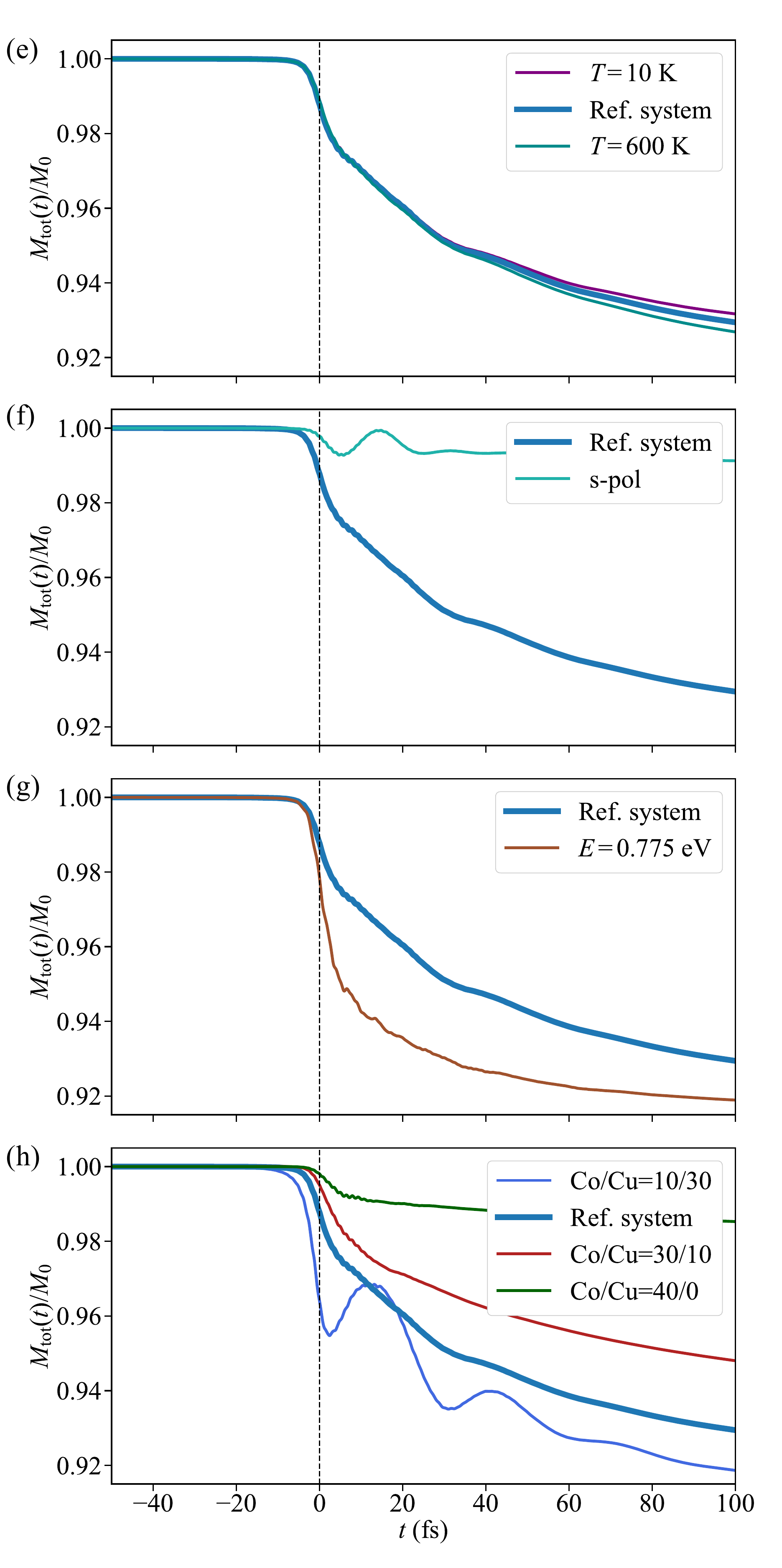}
	\caption{Comparing systems: magnetization of the Cu region $M_{\mathrm{Cu}}$ [left column, panels (a)--(d)] and of the entire sample $M_{\mathrm{tot}}$ [right column, panels (e)--(h)]. Each spectrum is normalized with respect to the initial total magnetization $M_{0}\equiv M_{\mathrm{tot}}(t_0) = M_{\mathrm{Co}}(t_0) + M_{\mathrm{Cu}}(t_0)$ at $t_0 = -\unit[50]{fs}$ of the associated system.}
	\label{fig:comparison-tot-Cu-magnetization}
\end{figure*}

In the main text we analyzed how variations of the setup parameters [cf.\ Table~1] affect the demagnetization in the Co region of the sample [cf.\ Fig.~6 in the main text as well as panels~(a) and~(c) in Fig.~\ref{fig:comparison-M_Co_N_exc-varying-temp_E_ph}]. For the sake of completeness, Fig.~\ref{fig:comparison-tot-Cu-magnetization} displays the magnetization of the respective Cu regions [left column, panels~(a)--(d)] and of entire samples [right column, panels~(e)--(h)]. 

As found for the measures and the observables the variations of light polarization and sample composition have a pronounced effect on the magnetization of both the Cu region and the entire sample. The influence of temperature is again weak. However, in contrast to the demagnetization in Co, the magnetization in the Cu region is not independent of the photon energy: the setup with a photon energy of $\unit[0.775]{eV}$, which is half as large as the one in the reference system, yields a magnetization that is approximately half as large as that of the reference system.

\subsection{Total magnetization and number of photo-excited electrons: variation of temperature and photon energy}
\label{app:M_Co_N_exc-varying-temp_E_ph}
\begin{figure}[h]
	\centering
	\includegraphics[width=\columnwidth]{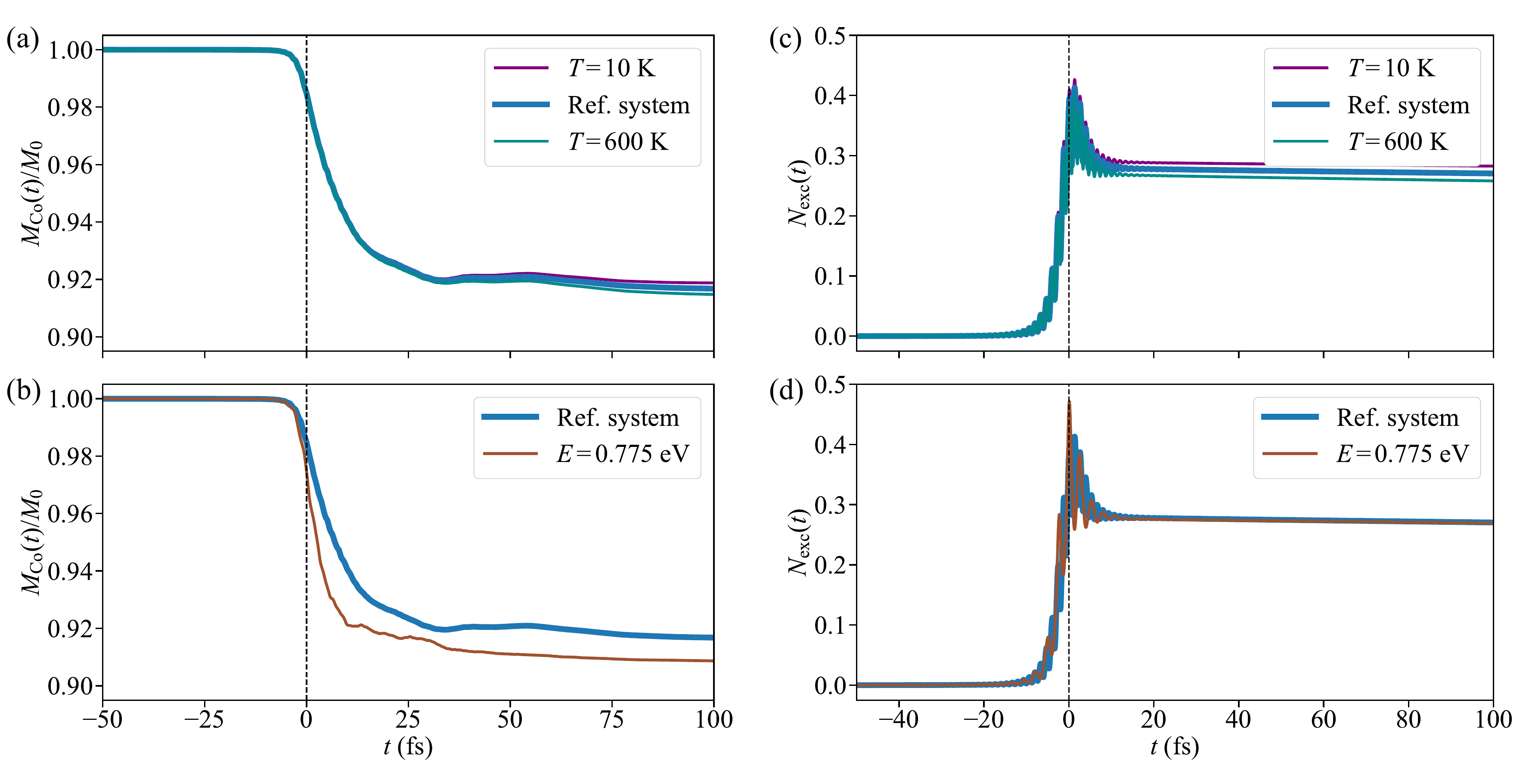}
	\caption{Comparing systems: magnetization of the entire sample ($M_{\mathrm{tot}}$, left) and the number of excited electrons ($N_{\mathrm{exc}}$, right). Each magnetization spectrum is normalized with respect to the magnetization $M_{0}$ at $t = -\unit[50]{fs}$ of the associated system. The variations of both temperature and photon energy yield minor effect on the selected observables.}
	\label{fig:comparison-M_Co_N_exc-varying-temp_E_ph}
\end{figure}

As for the measures, temperature has a minute effect on both the magnetization and the number of excited electrons [cf.\ Fig.~\ref{fig:comparison-M_Co_N_exc-varying-temp_E_ph}(a) and~(c)]. A higher temperature yields less initial magnetization because more electronic states with dominantly minority-spin orientation are occupied. The relative change, however, is almost independent of $T$. We note in passing that thermal fluctuations of the local magnetic moments, i.\,e.\ a noncollinear magnetic configuration, are not considered: the configuration is collinear and the temperature enters only via the occupation probabilities. Due to the modified initial occupation less electrons are excited after the laser pulse at a higher temperature.
The variation of the photon energy produces small changes in the Co magnetization and has no visible effect on the number of laser-excited electrons [panels (b) and (d)].

\end{document}